\documentclass[prl,floatfix,twocolumn,showpacs,amsmath,superscriptaddress]{revtex4-2}
\usepackage{hyperref}
\usepackage{graphicx}
\usepackage{braket}
\usepackage{color,xcolor}
\usepackage{bm}
\usepackage{array}
\usepackage{xcolor}
\usepackage{amsfonts}
\usepackage{amssymb}
\usepackage{bbm}
\usepackage{soul} 
\usepackage{multirow}
\newcommand{\tabincell}[2]{\begin{tabular}{@{}#1@{}}#2\end{tabular}}
\makeatletter

\newcommand{\Rmnum}[1]{\expandafter\@slowromancap\romannumeral #1@}
\newcommand{\ua}{\uparrow} 
\newcommand{\da}{\downarrow}
\newcommand{\ra}{\rightarrow}

\renewcommand{\vec}[1]{\mathbf{#1}}
\newcommand{\vk}{{\vec{k}}}

\begin{document}
\title{Quantum-geometry-induced intrinsic optical anomaly in multiorbital superconductors}

\author{Weipeng Chen}
\address{Shenzhen Institute for Quantum Science and Engineering, Southern University of Science and Technology, Shenzhen 518055, Guangdong, China}
\address{International Quantum Academy (SIQA), and Shenzhen Branch, Hefei National Laboratory, Futian District 518048, Shenzhen, China}
\address{Guangdong Provincial Key Laboratory of Quantum Science and Engineering, Southern University of Science
	and Technology, Shenzhen 518055, Guangdong, China}
\author{Wen Huang}
\email{huangw3@sustech.edu.cn}
\address{Shenzhen Institute for Quantum Science and Engineering, Southern University of Science and Technology, Shenzhen 518055, Guangdong, China}
\address{International Quantum Academy (SIQA), and Shenzhen Branch, Hefei National Laboratory, Futian District 518048, Shenzhen, China}
\address{Guangdong Provincial Key Laboratory of Quantum Science and Engineering, Southern University of Science
	and Technology, Shenzhen 518055, Guangdong, China}
\date{\today}

\begin{abstract}
	Bloch electrons in multiorbital systems carry nontrivial quantum geometric information characteristic of their orbital composition as a function of their wavevector. When such electrons form Cooper pairs, the resultant superconducting state naturally inherits aspects of the quantum geometry. In this paper, we study how this geometric character is revealed in the intrinsic optical response of the superconducting state. In particular, due to the superconducting gap opening, interband optical transitions involving states around the Fermi level are forbidden. This generally leads to an anomalous suppression of the optical conductivity at frequencies matching the band separation --- which could be significantly higher than the superconducting gap energy. We discuss how the predicted anomaly may have already emerged in two earlier measurements on an iron-based superconductor. When interband Cooper pairing is present, intraband optical transitions may be allowed and finite conductivity emerges at low frequencies right above the gap edge. These conductivities depend crucially on the off-diagonal elements of the velocity matrix in the band representation, i.e. the interband velocity --- which is related to the non-Abelian Berry connection of the Bloch states. 
\end{abstract}
\maketitle

{\bf Introduction.--}In a multiorbital system, Bloch electrons on the individual energy bands are generally not featureless particles. They exhibit nontrivial internal structure, in the sense that they possess quantum geometric information associated with their varying orbital composition as a function of the wavevector. This property manifests in the quantum transport: the electrical currents carried by the Bloch electrons in such multiorbital setting, having their origin in the physical motion of the underlying orbitals, are not determined solely by the group velocities of the individual Bloch states.

Let's illustrate the idea using the example of a general two-orbital model. We consider the two orbital degrees of freedom stemming from either two different electron orbitals residing on a monatomic lattice, or orbitals of the same symmetry each occupying one sublattice of a bipartite lattice. The kinetic Hamiltonian is given by $H= \sum_\vk \psi_{\vk}^\dagger \hat{H}^\text{orb}_\vk \psi_\vk$, where $\psi^\dagger_\vk = (a^\dagger_{\vk},b^\dagger_{\vk})$ in which $a^\dagger_{\vk}$ and $b^\dagger_{\vk}$ stand for the creation operators of the respective $a$ and $b$ orbitals, and $\hat{H}^\text{orb}_\vk$ generally takes the form, 
\begin{equation}
	\hat{H}^\text{orb}_\vk = \begin{pmatrix}
		\xi_{a,\vk}  &  \lambda_\vk \\
		\lambda_\vk^\ast & \xi_{b,\vk} 
	\end{pmatrix} \,.
	\label{eq:H0}
\end{equation}
Here, $\xi_{a(b),\vk}$ represent the dispersion relation of the two orbitals in the unhybridized limit and $\lambda_\vk$ describes the orbital mixing. For brevity, we have ignored spin-orbit coupling (SOC) for now and suppressed the spin indices. The forms of $\lambda_\vk$ and $\xi_{\vk}=\xi_{a,\vk}-\xi_{b,\vk}$ are essential for characterizing the quantum geometry (QG) of the resultant Bloch electrons. Since electrical current is generated by the hopping of the orbitals on the lattice, it is connected to the Hamiltonian (\ref{eq:H0}) via the relation $J_{\mu,\vk} =e \psi_\vk^\dagger \hat{V}^\text{orb}_{\mu,\vk} \psi_\vk$, where $\hat{V}^\text{orb}_{\mu,\vk}= \partial_{k_\mu}\hat{H}^\text{orb}_\vk$ is the velocity operator. Formally, the operator is obtained from the standard Peierls substitution by changing $\vk$ to $\vk-e \vec{A}$ in the kinetic Hamiltonian and then taking $\partial \hat{H}^\text{orb}_{\vk-e\vec{A}}/\partial \vec{A}|_{\vec{A}\ra 0}$. This operator can be rewritten in the (Bloch) band basis through a unitary transformation. We write $J_{\mu,\vk} =e \phi_\vk^\dagger \hat{V}_{\mu,\vk} \phi_\vk$, where $\phi_\vk^\dagger=(c_{1,\vk}^\dagger,c_{2,\vk}^\dagger)$ with $1,2$ denoting the band indices, and 
\begin{equation}
	\hat{V}_{\mu,\vk} = \hat{\mathcal{V}}^{-1}_\vk \partial_{k_\mu}\hat{H}^\text{orb}_\vk \hat{\mathcal{V}}_\vk = \begin{pmatrix}
		V_{\mu,\vk}^{11} & V_{\mu,\vk}^{12} \\
		V_{\mu,\vk}^{21} & V_{\mu,\vk}^{22}
	\end{pmatrix} \,,
	\label{eq:Vop}
\end{equation}
where $\hat{\mathcal{V}}_\vk$ is a unitary matrix that diagonalizes $\hat{H}^\text{orb}_\vk$. The diagonal elements of (\ref{eq:Vop}), i.e. the intraband velocities $V^{nn}_{\mu,\vk}$, can be shown to simply equal the band group velocities. The interband velocity, on the other hand, is given by, 
\begin{eqnarray}
	V_{\mu,\vk}^{mn} &=&\langle m,\vk|\partial_{k_\mu} \hat{H}^\text{orb}_\vk|n,\vk \rangle \nonumber \\
	&=&(\epsilon_{m,\vk}-\epsilon_{n,\vk}) \langle m,\vk|\partial_{k_\mu} |n,\vk \rangle\,,~~~(m\neq n)
\end{eqnarray}
where $\epsilon_{m,\vk}$ is the energy dispersion of the $m$-th band with eigenvector $|m,\vk \rangle$. The object $i\langle m,\vk|\partial_{k_\mu} |n,\vk \rangle$ is known as the non-Abelian Berry connection and is involved in the definition of the quantum geometric tensor~\cite{Provost:80,Liang:17,Iskin:19,Li:20,Topp:21,Ahn:21}. It is finite provided the orbital composition of the Bloch states, i.e. the relative amplitude and/or relative phase of the orbitals involved, vary continuously in momentum space. By inspection, this is equivalent to stating that: 1), the orbital mixing $\lambda_\vk$ is finite and 2), at least one of $\xi_{\vk}$ and $\lambda_\vk$ is a varying function of $\vk$. The interband velocity showcases a profound consequence of the QG, as it depicts how electrons from distinct energy bands still `talk' to each other as far as charge transport is concerned. 

In a weak-coupling superconducting state, electrons on the Fermi level most actively participate in Cooper pairing. Without considering interband pairing, the fermionic excitation spectra of the two bands are decoupled. However, by the same reasoning stated above, the Bogoliubov quasiparticles of one band shall still inherently connect to the unpaired electrons of the other via the QG. It is thus natural to expect this quantum connection to reveal itself in the superconducting electromagnetic response. There were previous studies along similar direction about the band geometric effect on the superfluid stiffness in flatband systems~\cite{Peotta:15,Julku:16,Liang:17}, and applications to theories of the twisted bilayer graphene followed recently~\cite{Hu:19,Julku:20,Xie:20,Hazra:19,Verma:21}. More of its unusual aspects await investigation. In this paper, we demonstrate how the geometric character of the superconducting electrons influences the intrinsic optical conductivity. 

\begin{figure}
	\includegraphics[width=9.cm]{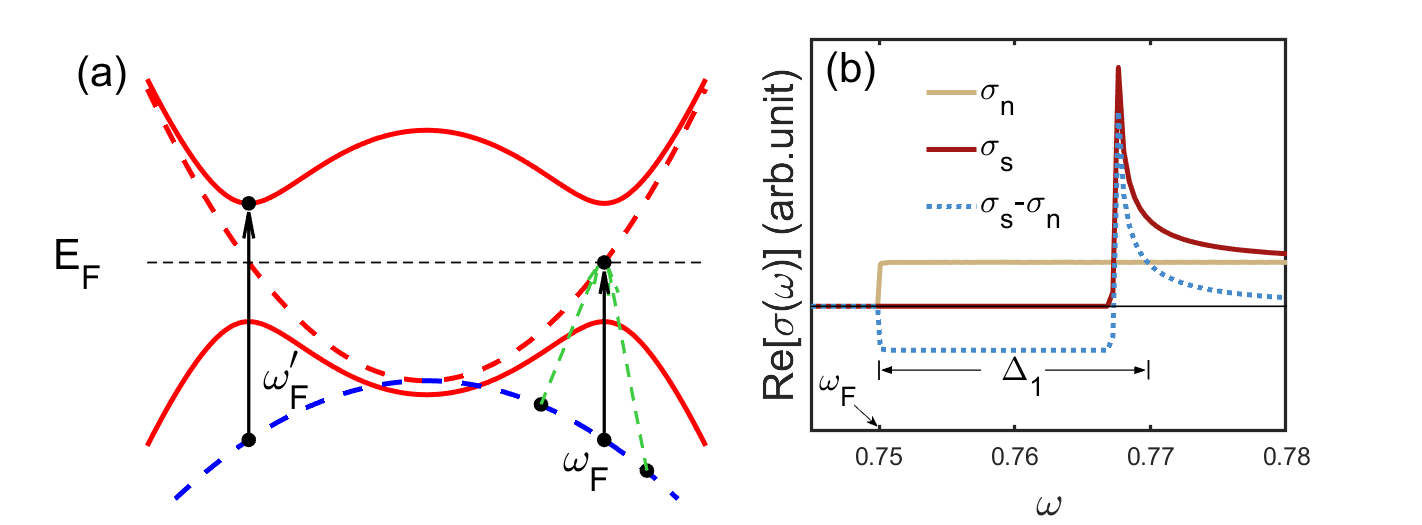}
	\caption{(color online) (a) Schematic interband optical transitions (arrows) in a $d_{xz}$-$d_{yz}$ model with only one band crossing the Fermi level. Dashed curves sketch the normal state band structure, and the red solid curves the Bogoliubov quasiparticle spectra of the superconducting band. The black arrows denote transitions that contribute to the intrinsic optical transitions in the normal (dotted) and superconducting (solid) states, while the green dotted arrows exemplify momentum-non-conserving transitions induced by disorder scattering in normal state. (b) The real part of the optical conductivity of normal and superconducting phases near the `cutoff' frequency $\omega_F \approx 0.75$. The dotted curve shows their difference, which exhibits a dip between $\omega_F$ and roughly $\omega_F+\Delta_1$. In the calculation, we took $\xi_{a(b),\vk} =\frac{k_x^2}{2m_{\parallel(\perp)}} + \frac{k_y^2}{2m_{\perp(\parallel)}} -\mu $ and $\lambda_\vk= \lambda k_x k_y$, with $(m_\parallel, m_\perp, \mu, \lambda)=(1, -2, 0.5, -0.75)$. Superconducting pairing is assumed to take place only on the band that crosses the Fermi level, with pairing gap $\Delta_1=0.02$. }
	\label{fig01}
\end{figure}

{\bf Formulation.--} In what follows, we will consistently formulate our theory in the band representation, for both normal and superconducting states. In this basis, the eigenvectors of the individual bands acquire the simple forms $(1,0)^T$ and $(0,1)^T$, for which we continue to use the designation $|m(n),\vk\rangle$. Let's first consider the normal state response. Within linear-response theory, the conductivity is given by the following current-current correlator,
\begin{equation}
	\sigma_n(\omega) =\frac{T}{\omega} \sum_{\vk,\omega_n} \text{Tr}\left[ \hat{V}_{x,\vk} \hat{G}_\vk(\omega_n)\hat{V}_{x,\vk} \hat{G}_\vk(\omega_n+\omega) \right] \,.
	\label{eq:SigmaNormal}
\end{equation}
We shall focus on the real part of $\sigma(\omega)$. Expressing the Greens' function in the spectral representation, one obtains, for the real part that we shall focus on, 
\begin{equation}
	\text{Re}[\sigma_n(\omega) ]= \sum_{m,n,\vk} \left|\langle m,\vk| \hat{V}_{x,\vk}  | n,\vk \rangle \right|^2 F(\omega;\epsilon_{m,\vk},\epsilon_{n,\vk})
	\label{eq:ReSigmaN}
\end{equation}
We have adopted a shorthand expression $F(\omega;x_1,x_2)=[f(x_1)-f(x_2)]\delta(\omega + x_1-x_2)/\omega$, where $f(x)$ stands for the Fermi distribution function. At zero temperature, it has finite value only at wavevectors satisfying $\epsilon_{m,\vk}<0<\epsilon_{n,\vk}$, and is therefore cut off in momentum space by the Fermi surface(s). The corresponding `cutoff' frequencies $\omega_\text{F}$ (Fig.~\ref{fig01}), which measure the band separation at the Fermi wavevectors, are important quantities to keep in mind. 

Obviously, only momentum-conserving interband transition (i.e. $m\neq n$) contributes, with $\langle 2,\vk| \hat{V}_{x,\vk}  | 1,\vk \rangle=V_{x,\vk}^{21}$. Hence the interband velocity fully characterizes the geometric footprints in the intrinsic conductivity. Note that simply inserting identity operator $\hat{\mathcal{V}}_\vk^{-1}\hat{\mathcal{V}}_\vk$ in between the $\hat{V}_x$'s and $\hat{G}$'s in the first line of (\ref{eq:SigmaNormal}) restores the expression to the orbital basis formulation, and the same result will follow. 

Turning to the superconducting state, we consider for simplicity a scenario where only one of the bands (say, band-1) crosses the Fermi energy and superconducts with a simple constant $s$-wave gap, while the other sits below the Fermi energy (Fig.~\ref{fig01}(a)). The general Bogoliubov-de Gennes (BdG) Hamiltonian reads, 
\begin{equation}
	H_\text{BdG} =\sum_{\vk, n, s}\epsilon_{n,\vk} c^\dagger_{n,\vk s}c_{n,\vk s} + \sum_{\vk}\Delta_{1,\vk} c_{1,\bar{\vk}\ua}c_{1,\vk\da} +h.c. .
	\label{eq:HBdG}
\end{equation}
In the subblock associated with the Nambu basis $(c_{1,\vk\ua},c_{2,\vk\ua},c^\dagger_{1,\bar{\vk}\da},c^\dagger_{2,\bar{\vk}\da})^T$ where $\bar{\vk}=-\vk$, the velocity operator becomes,
\begin{eqnarray}
	\hat{\widetilde{V}}_{x,\vk}&=&\begin{pmatrix}
		\hat{\mathcal{V}}^{-1}_\vk \partial_{k_x} \hat{H}_\vk \hat{\mathcal{V}}_\vk & \\
		& \hat{\mathcal{V}}^{*-1}_{\bar{\vk}} \partial_{k_x} \hat{H}^*_{\bar{\vk}} \hat{\mathcal{V}}^*_{\bar{\vk}} 
	\end{pmatrix}\nonumber \\
	&=&\begin{pmatrix}
		\hat{V}_{x,\vk} & \\
		& -\hat{V}^*_{x,\bar{\vk}} 
	\end{pmatrix}.
\end{eqnarray}
Generalizing (\ref{eq:SigmaNormal}) to the superconducting state, one arrives at an expression for $\text{Re} \left[\sigma_s(w)\right]$ similar to (\ref{eq:ReSigmaN}), but with $\epsilon_{m(n),\vk}$ and $|m(n),\vk\rangle$ replaced respectively by the Bogoliubov quasiparticle dispersion and eigenvectors. 

As in the normal state, virtual excitations that contribute to the conductivity involve interband processes. A representative transition is indicated by the solid black arrow in Fig.~\ref{fig01} between a negative-energy state associated with band-2, $|\bar{2}, \vk \rangle_s = (0,1,0,0)^T$, and a positive-energy state associated with band-1, $|1,\vk \rangle_s = (\Delta_{1,\vk},0,E_{1,\vk}-\epsilon_{1,\vk},0)^T/\mathcal{N}_{1,\vk}$ where $E_{1,\vk} = \sqrt{\epsilon_{1,\vk}^2+\Delta_{1,\vk}^2}$ and $\mathcal{N}_{1,\vk}=\sqrt{2E_{1,\vk}(E_{1,\vk}-\epsilon_{1,\vk})}$ is a normalization factor. Note that we use the subscript `s' to designate states in the superconducting state. One then obtains a transition matrix element,
\begin{equation}
	_s\langle \bar{2},\vk| \hat{\widetilde{V}}_{x,\vk}  | 1,\vk \rangle_s=\frac{\Delta_{1,\vk}}{\mathcal{N}_{1,\vk}} V_{x,\vk}^{21} \,.
\end{equation}

Another process involving their counterpart particle-hole symmetric states $|2,\bar{\vk}\rangle_s$ and $|\bar{1},\bar{\vk}\rangle_s$ leads to the same expression. Put together, and with a proper account of the double-counting in Nambu representation, the full expression for the real part of conductivity becomes,
\begin{equation}
	\text{Re} [\sigma_s(w)] =\sum_{\vk} \left|\frac{\Delta_{1,\vk}}{\mathcal{N}_{1,\vk}} V_{x,\vk}^{21} \right|^2 F(\omega;-E_{2,\vk},E_{1,\vk})
\end{equation}
where $E_{2,\vk}=|\epsilon_{2,\vk}|$ for an unpaired band-2. Due to the superconducting gap opening in band-1, transitions involving states in the immediacy of the Fermi surface are forbidden. As a consequence, the conductivity in the superconducting state shall show an anomalous suppression within a width of $\Delta_1$ around the `cutoff' frequencies defined above. On the other hand, a pile-up of density of states at the continuum edge shall give rise to a peak right above $\omega_F^\prime \approx \omega_F+\Delta_1$. This is confirmed in Fig.~\ref{fig01} using a continuum two-orbital model with $d_{xz}$ and $d_{yz}$ orbitals. 

\begin{figure}
	\includegraphics[width=9.cm]{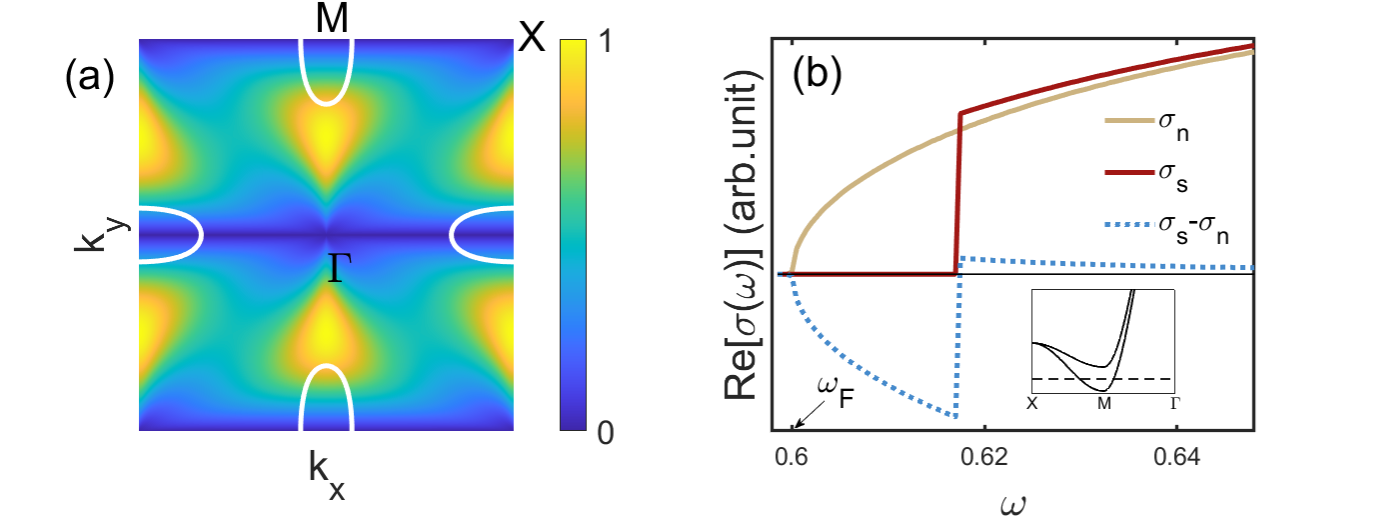}
	\caption{(color online) (a) Fermi surface (white contour) and normalized magnitude of the interband velocity $|V^{12}_{x,\vk}|$ in a square lattice $d_{xz}$-$d_{yz}$ model. (b) The real part of the optical conductivity in the normal and superconducting states, and their difference. The model has $\xi_{a(b),\vk}=-2t_{\parallel(\perp)}\cos k_x-2t_{\perp(\parallel)}\cos k_y-4t'\cos k_x\cos k_y-\mu$ and $\lambda_{ab,\vk}=\lambda \sin k_x \sin k_y$, where $(t_{\parallel}, t_{\perp}, t^\prime,\lambda,\mu)=(1,0.8,-0.6,0.4,-2.4)$. {\bf Inset}: Band structure near the M-point. Only one band crosses Fermi level with a pairing gap $\Delta_1=0.02$.}
	\label{fig02}
\end{figure}

In lattice models with band anisotropy, the peak structure could be smeared out. We now generalize the above analysis to a square lattice model. In a representative calculation shown in Fig.~\ref{fig02}, our model produces two Fermi pockets at the M points of the BZ --- reminiscent of the scenario in some iron-based superconductors~\cite{Fn1}. Figure~\ref{fig02}(a) depicts the momentum-space distribution of the absolute value of the x-component of the interband velocity. Notably, unlike the intraband velocities, it does not vanish at $k_x=0, \pi$. As shown in Fig~\ref{fig02}(b), $\text{Re}[\sigma_s(\omega)-\sigma_n(\omega)]$ no longer exhibits the peak structure, whereas the dip above $\omega_F$ persists. 

Besides anisotropy, real materials always contain finite amount of disorder scatterings which enable momentum-non-conserving interband transitions. As a consequence, the gap opening prohibits not only the `direct' optical transitions at $\omega_F$, but also some `indirect' transitions as exemplified by the green dashed arrows in Fig.~\ref{fig01} (a). Hence the anomalous suppression is not necessarily restricted to a narrow width of $\Delta_1$ above $\omega_F$, but could potentially take place at all frequencies associated with indirect band separations, which amount to a frequency span as wide as the bandwidth of the unpaired band. Interestingly, one such optical anomaly was indeed detected in an earlier measurement on the iron-based superconducting compound Ba$_{0.68}$K$_{0.32}$Fe$_{2}$As$_2$~\cite{Charnukha:10}. There, the superconductivity-induced conductivity suppression centers around $\omega=2.5$ eV, and spans a broad frequency range of 1 eV. The effect was ascribed to interband transitions involving the strongly hybridized Fe-d and As-p orbitals. 

A more pronounced anomaly was reported in the same material at lower frequencies above the gap edge, and was attributed to spin-fluctuation–assisted scattering effects among the multiple bands dominated by the Fe-d orbitals near the Fermi energy~\cite{CharnukhaPRB}. However, as we describe in detail in the Supplementary~\cite{Supp}, our theory is also able to produce a qualitatively similar anomalous suppression, using a model that captures some essential features of the band and gap structure of this material around the $\Gamma$-point~\cite{Ding:2011}. 

\begin{figure}
	\includegraphics[width=9.cm]{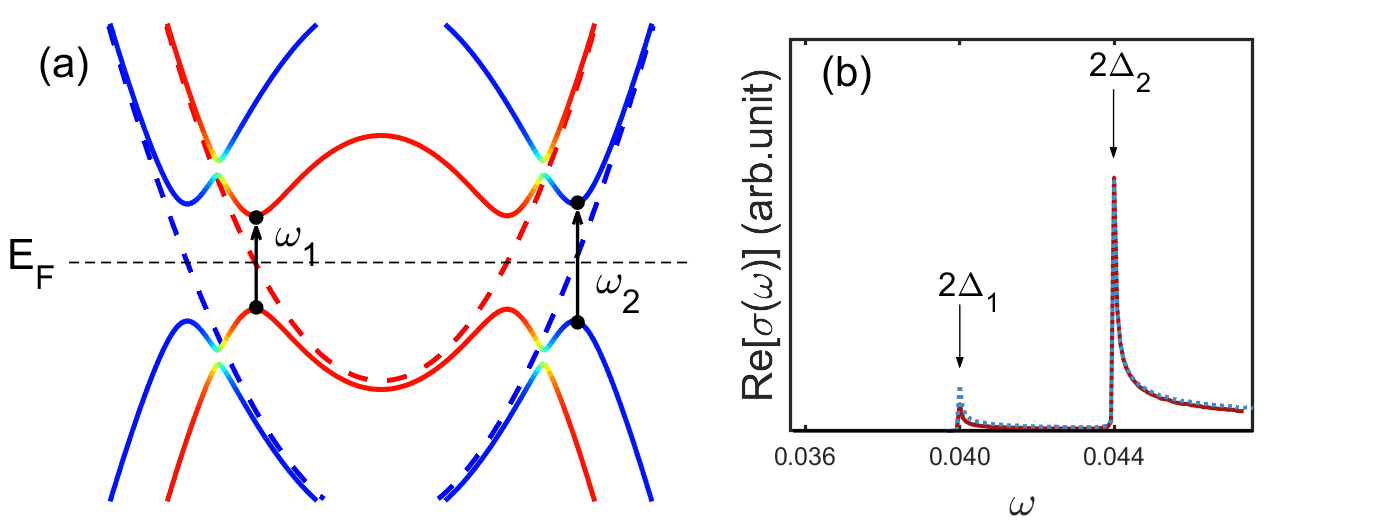}
	\caption{(color online) (a) Schematic intraband optical transitions (black arrows) induced by interband Cooper pairing. The dashed and solid curves depict the energy dispersion in the normal and superconducting states of a continuum $d_{xz}$-$d_{yz}$ model with finite SOC. (b) Real part of the low frequency optical conductivity induced by interband pairing. The solid (red) and dotted (blue) curves show results obtained from the numerical calculation and the perturbative expansion, respectively. The continuum Hamiltonian with SOC is given in the Supplementary \cite{Supp} and we took $(m_\parallel, m_\perp, \mu, \lambda, \eta)=(1, 0.2, 0.8, 0.5, 0.3)$. The pairing amplitudes are $(\Delta_1, \Delta_2, \Delta_{12})=(0.02, 0.022, 0.005)$ and the constant $\gamma=1$.}
	\label{fig03}
\end{figure}

{\bf Interband pairing induced conductivity.--} It must be stressed that, in the absence of interband Cooper pairing, low-energy intraband transitions (Fig.\ref{fig03}(a)) are absent in the clean limit. This explains the vanishing of $\text{Re}[\sigma_s(\omega)]$ at low frequencies. It is also reminiscent of the absence of longitudinal excitations in single-band superconductors under the perturbation of a weak uniform magnetic field~\cite{Schrieffer:18}, i.e. $H^\prime = \int d\vec{r}~e\hat{\widetilde{\boldsymbol{V}}}\cdot\vec{A}$, which underlies the Meissner effect therein. One way to see the connection is by recognizing, in single-band systems, the vanishing of $_s\langle \bar{m},\vk | \hat{\widetilde{V}}_{\mu,\vk} |m,\vk\rangle_s$, where $\hat{\widetilde{V}}_{\mu,\vk}$ is proportional to an identity matrix in this case.

One natural question is how interband pairing affects the low-frequency electromagnetic response. As we focus on translation invariant models, we only consider interband Cooper pairs formed between electrons of opposite wavevectors. Such interband pairing is customarily ignored in many literature, however, there exists no symmetry constraint to forbid it in realistic multiband superconductors. In fact, it is particularly relevant for systems where the band separation is small, and more so for superconducting pairings driven by electron-electron correlations -- where the Coulomb-derived effective attractive interactions are not necessarily confined to a thin layer of phase space around the Fermi surface~\cite{Fn2}. 

We illustrate the idea in the continuum limit of the square lattice $d_{xz}$-$d_{yz}$ model. For generality purpose and to avoid ambiguity at $\vec{k}=0$ due to band degeneracy, we add a finite SOC. We further assume that both bands cross the Fermi level and superconduct, and that they are proximate in energy, making it more reasonable to speak about sizable interband pairing in the system. 

Around $\vec{k}=0$, the two bands are classified according to their total angular momentum eigenvalues $J=L+S$, designated as follows,
\begin{eqnarray}
	\text{Band-1:~~~~~~}&&  |j_z=\frac{3}{2}\rangle = |d_{xz} +i d_{yz}\rangle\otimes |\ua \rangle, \nonumber \\
	&&  |j_z=-\frac{3}{2}\rangle = |d_{xz} -i d_{yz}\rangle\otimes |\da \rangle; \nonumber \\
	&& \nonumber \\
	\text{Band-2:~~~~~~}&&  |j_z=-\frac{1}{2}\rangle = |d_{xz} -i d_{yz}\rangle\otimes |\ua \rangle, \nonumber \\
	&&  |j_z=\frac{1}{2}\rangle = |d_{xz}+i d_{yz}\rangle\otimes |\da \rangle\,. \nonumber
\end{eqnarray}
The symmetry properties of these states directly affect the forms of the interband pairings, which requires a special attention as they in general do not share the same basis functions as that of the intraband pairing~\cite{Samokhin:20}. This can be understood by noting that, when acting on an interband pairing $\Delta_{12,\vec{k}}c_{1,\bar{\vec{k}}}c_{2,\vec{k}}$, any symmetry operation must simultaneously transform $\Delta_{12,\vec{k}}$, as well as the wavefunctions of the two constituent electrons which exhibit distinct symmetries. 

We take as an example a superconducting state with $A_{1g}$ symmetry in the $D_{4h}$ point group, where the intraband pairings on the two bands acquire usual $s$-wave forms. As is analyzed more systematically in the Supplementary~\cite{Supp,Huang:19}, the interband pairing develops only between $j_z=\frac{3}{2}$ and $j_z=\frac{1}{2}$ states, and between $j_z=-\frac{3}{2}$ and $j_z=-\frac{1}{2}$ states, and it shall acquire a general form of $(k^2_x-k^2_y)+i \gamma k_xk_y$, where $\gamma$ is an inessential real constant that depends on microscopic details. Such a pairing combines with the symmetries of the constituent electrons to yield an overall $A_{1g}$ symmetry. Consider, for instance, a $C_4$ rotation. It changes $\Delta_{12,\vk}$ to $-\Delta_{12,\vk}$, $|j_z=\frac{3}{2}\rangle$ to $e^{i\frac{3\pi}{4}}|j_z=\frac{3}{2}\rangle$, and $|j_z=\frac{1}{2}\rangle$ to $e^{i\frac{\pi}{4}}|j_z=\frac{1}{2}\rangle$. Hence, the interband pairing is invariant under $C_4$, and the same holds under other symmetry operations. 

Treated as a perturbation to the original BdG Hamiltonian, the interband pairing can be written, in the sub-basis $(c_{\frac{3}{2},\vk},c_{-\frac{1}{2},\vk},c^\dagger_{-\frac{3}{2},\bar{\vk}},c^\dagger_{\frac{1}{2},\bar{\vk}})^T$, as~\cite{Supp}
\begin{equation}
	\hat{H}_{\delta,\vk}=\begin{pmatrix}
		&&&\Delta_{12,\vk} \\
		&&\Delta_{21,\vk} &\\
		&\Delta^*_{21,\vk} &&\\
		\Delta^*_{12,\vk} &&&\\
	\end{pmatrix}\,,
\end{equation}
where $\Delta_{12,\vk}=\Delta^*_{21,\vk}= \Delta_{12}[(k^2_x-k^2_y)+i \gamma k_xk_y]$. By second-order perturbation theory, the intraband transition matrix element follows as,
\begin{eqnarray}
	&&_s\langle \bar{m}',\vk |\hat{\widetilde{V}}_{\mu,\vk} |m',\vk\rangle_s\nonumber \\
	&=&(\alpha_{I,\vk}-\alpha_{II,\vk}  ) 
	(V^{nm}_{\mu,\vk}u_mv_n-V^{mn}_{\mu,\vk}v_mu_n) \nonumber \\
	&+&(\beta_{I,\vk}-\beta_{II,\vk}  )(V^{nm}_{\mu,\vk}u_mu_n^*+V^{mn}_{\mu,\vk}v_mv_n^*),~~(m\neq n).~~~~~
	\label{eq:intraTransition}
\end{eqnarray}
Here, $|m^\prime,\vk\rangle_s$ and $|\bar{m}^\prime,\vk\rangle_s$ with $m=1,2$ denote the respective perturbed particle- and hole-like quasiparticle states associated with the corresponding bands, $ (u_m,v_m)=(\Delta_{m,\vk},E_{m,\vk}-\epsilon_{m,\vk})/\mathcal{N}_{m,\vk}$, and the remaining coefficients are linearly related to the interband pairing as follows,
\begin{eqnarray}
	\alpha_{I,\vk}&=&\frac{\Delta_{mn,\vk}^*u_mv_n^*+\Delta_{nm,\vk}v_mu_n^*}{E_{n,\vk}-E_{m,\vk}}, \nonumber \\
	\alpha_{II,\vk}&=&\frac{\Delta_{nm,\vk}^*u_mv_n^*+\Delta_{mn,\vk}v_mu_n^*}{E_{n,\vk}-E_{m,\vk}}, \nonumber \\
	\beta_{I,\vk}&=&\frac{-\Delta_{mn,\vk}^*u_mu_n+\Delta_{nm,\vk}v_mv_n}{E_{n,\vk}+E_{m,\vk}}, \nonumber \\
	\beta_{II,\vk}&=&\frac{ -\Delta_{nm,\vk}^*u_mu_n+\Delta_{mn,\vk}v_mv_n}{E_{n,\vk}+E_{m,\vk}}. 
\end{eqnarray}
The dependence on the interband velocity, and therefore on the QG, is evident in (\ref{eq:intraTransition}). Figure~\ref{fig03} (b) presents the numerically evaluated low-frequency optical conductivity for a model where the intraband pairings on the two bands are comparable and the interband pairing amplitude is roughly four times smaller. Also drawn is the result obtained from the perturbative analysis (blue dotted), which shows excellent agreement. The two prominent peaks correspond to the gap edge of the two respective bands. 

Ahn and Nagaosa~\cite{Nagaosa:21} have also studied the low-frequency intrinsic conductivity of similar origin. Approached from a pure orbital-basis perspective, they identified certain symmetry criterion that could enable intraband transitions. We checked that our model falls in the category of class DIII in their effective Altland–Zirnbauer symmetry classification, where intraband excitation are allowed when there exists finite SOC. Our study contributes two significant advances, one is to pinpoint interband pairing as an essential ingredient for generating low-frequency intrinsic conductivity, and the other is to emphasize the critical role of the QG. 

{\bf Conclusion.--} The quantum geometric character of the superconducting Bloch electrons in multiband systems has largely gone unnoticed until recently. In this paper, we demonstrated how it influences the intrinsic optical conductivity in a nontrivial fashion. We showed that, as interband transitions involving quasiparticles around the Fermi level are forbidden due to superconducting gap opening, the optical conductivity shall generally exhibit an anomalous suppression at frequencies matching the band separation. The presence of interband Cooper pairing may further enable low-energy intraband optical transitions, thereby inducing low-frequency conductivity in the clean limit. These intrinsic contributions to the conductivity are closely related to the interband velocity, which, by itself, is a manifestation of the quantum geometry of the Bloch electrons. We expect our theory to be applicable to a broad spectrum of superconductors which exhibit multiband or multiorbital character, and we pointed out the possible connection to previously reported optical conductivity anomalies in an iron-based superconductor. 

{\bf Note added.--} After our paper appeared on arXiv, we became aware of a similar work by Ahn and Nagaosa \cite{Ahn:2021}, which also studied the anomalous changes in the optical spectral weight at high frequencies due to Bloch quantum geometry. The authors further pointed out the potential relevance to the long-standing puzzle of spectral weight transfer reported in underdoped and optimally doped cuprate superconductors (e.g. Ref.~\cite{Molegraaf:2002}).  

{\bf Acknowledgments.-} We acknowledge helpful discussions with Junfeng Dai, Jiawei Mei, Nanlin Wang and Zhongbo Yan. We are also indebted to Alexander Boris for drawing our attention to Ref.~\cite{CharnukhaPRB} and for his comments. This work is supported by NSFC under grant No.~11904155 and the Guangdong Provincial Key Laboratory under Grant No.~2019B121203002. Computing resources are provided by the Center for Computational Science and Engineering at Southern University of Science and Technology.

\newpage

\newpage
\section{Supplementary materials}
\renewcommand\thefigure{S}
\setcounter{figure}{0}
\subsection{Lower-frequency anomaly in Ba$_{0.68}$K$_{0.32}$Fe$_{2}$As$_2$}

\begin{figure}
	\includegraphics[width=9cm]{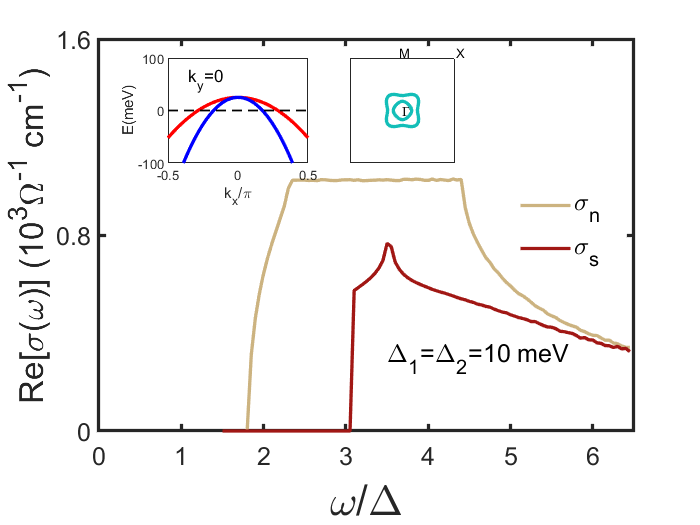}
	\caption{(color online) The simulated real part of the lower-frequency optical conductivity in Ba$_{0.68}$K$_{0.32}$Fe$_{2}$As$_2$ using a $d_{xz}$-$d_{yz}$ model, in the normal (light brown) and superconducting (dark red) states. The kinetic part of the Hamiltonian is given in Fig. 1 in the maintext and we take $(m_\parallel, m_\perp, \mu, \lambda)=(-4, -1.5, -0.1, 0.32)$ for the present calculation. The left inset shows the band structure around the $\Gamma$-point, while the right inset plots the shape of the Fermi surfaces. Superconducting gaps on the two bands are set at $\Delta_1=\Delta_2=10$ meV. To compare with the experimental data in Ref. \cite{ref2}, we have used the actual unit $\frac{e^2}{\pi \hbar l}$ in the calculation of the conductivity. Here, $l$ is the lattice constant of the iron-based superconductor, which is taken to be 3.9 \r{A}.}
	\label{figSupp}
\end{figure}

In this section, we construct an effective two-orbital $d_{xz}$-$d_{yz}$ model to simulate the anomalous suppression of the optical conductivity in the superconducting state of Ba$_{0.68}$K$_{0.32}$Fe$_{2}$As$_2$ and make a comparison to the experimental report~\cite{ref2}. For a representative set of parameters, our model produces a band structure shown in the insets of Fig.~\ref{figSupp}, with two hole-like bands crossing the Fermi energy around the $\Gamma$-point. The actual band structure and the Fermi surface shapes are more complicate, however, this simplified model serves our illustrative purpose. Consistent with the implications from angle-resolved photoemission studies (e.g. Ref.~\onlinecite{ref1}), the band top is set at $\sim$ 20 meV above the Fermi level, and the band separation in the region between the two Fermi surfaces ranges from $\sim$ 10 to 50 meV. These energy scales are comparable to the superconducting gap size at roughly $\sim$ 10 meV~\cite{ref2}. This would have considerable effect on the detailed behavior of the superconductivity-induced suppression. Assuming a constant isotropic gap, the calculation produces normal and superconducting conductivities as shown in Fig.~\ref{figSupp}. As one can see, besides the strongest suppression seen within a frequency span of the width of the gap size, a weaker but still noticeable suppression extends to much higher frequencies. This agrees qualitatively with the experimental observation in Ref.~\onlinecite{ref2}. 

Note that this calculation has only accounted for the conductivity generated by interband optical transitions. The intraband optical transitions enabled by interband pairing, which is not included in the present calculation, does not lead to any qualitative difference. 

\renewcommand{\theequation}{S\arabic{equation}}
\setcounter{equation}{0}
\subsection{$A_{1g}$ gap functions in the band representation}
Following Ref.~\cite{ref3}, the normal-state Hamiltonian of the spin-orbit-coupled $d_{xz}$-$d_{yz}$ model on a square lattice reads, 
\begin{eqnarray}\hat{H}^\text{orb}_\vk=\frac{\xi_{a,\vk}+\xi_{b,\vk}}{2}+\frac{\xi_{a,\vk}-\xi_{b,\vk}}{2} \sigma_z +\lambda_{\vk}\sigma_x+\eta s_z \sigma_y, \nonumber \\
\end{eqnarray}
where $s_i$ and $\sigma_i$ are Pauli matrices acting on spin and orbital spaces and $\eta$ represents the strength of SOC. This model has $D_{4h}$ point group symmetry and in two-dimensions, all of the orbital-basis superconducting gap functions belonging to the $A_{1g}$ irreducible representation are listed in Table \ref{tab1}~\cite{ref3}. One can see that the Cooper pairing only takes place between antiparallel spins in this model. To obtain the pairing in the band basis, we utilize the following unitary transformation:
\begin{eqnarray}
	\hat{\Delta}_\vk&=&\hat{\mathcal{V}}^{-1}_\vk \hat{\Delta}^\text{orb}_\vk \hat{\mathcal{V}}^*_{\bar{\vk}},
\end{eqnarray}
where $\hat{\mathcal{V}}_\vk$, as defined in the maintext, diagonalizes $\hat{H}^\text{orb}_\vk$. 

The band-basis gap functions contains both intraband and interband parts. As expected, the intraband pairing derived from different $\hat{\Delta}^\text{orb}_\vk$ exclusively acquire $s$-wave forms. The interband pairing, on the other hand, varies among the multiple $\hat{\Delta}^\text{orb}_\vk$, as given in Table~\ref{tab1}. Due to orbital hybridization and SOC, the multiple forms of $\hat{\Delta}^\text{orb}_\vk$ in the same symmetry channel may simultaneously appear~\cite{ref3}. As a consequence, the different forms of interband pairings shall naturally mix, giving rise to a generic form,
\begin{equation}
	\hat{\Delta}_\vk \propto  (k^2_x-k^2_y) \sigma^\prime_x + \gamma k_xk_y \sigma_y^\prime\otimes \bm{z}\cdot \bm{s} \,,
\end{equation}
where $\sigma^\prime$ acts in the band manifold, and $\gamma$ is a real constant that is determined by inessential microscopic details. It has the structure of a $d+id$ pairing in an ordinary single-orbital/band model. One can see that interband pairing in the present case occurs only between antiparallel spins, which can be mapped to electron pairs with $j_z=\frac{3}{2}$ and $j_z=\frac{1}{2}$, and those with $j_z=-\frac{3}{2}$ and $j_z=-\frac{1}{2}$.
\renewcommand{\thetable}{S\Roman{table}}
\begin{widetext}
	\renewcommand\arraystretch{1.3}
	\begin{table*}
		\caption{$A_{1g}$ superconducting basis functions in the $d_{xz}$-$d_{yz}$ model on a square lattice. Pauli matrices $\sigma_i$, $\sigma'_i$ and $s_i$ operate in orbital, band and spin space, respectively. Note that we only keep the leading-order terms up to the quadratic order in $k$. $\alpha$ is an inessential real coefficient. The actual pairing gap functions are obtained by multiplying the basis function by $is_y$. }
		\begin{tabular}{p{2cm}<{\centering}|p{2cm}<{\centering}|p{3cm}<{\centering}|p{5cm}<{\centering}|p{5cm}<{\centering}}
			\hline
			\hline
			$\Delta_\vk^\text{orb}$&$\bm{1}$&$(k^2_{x}-k^2_y)\sigma_z$&$k_xk_y\sigma_x$&$\sigma_y\otimes \bm{z}\cdot \bm{s}$\\
			\hline
			\tabincell{c}{interband \\ pairing}&0&$(k^2_x-k^2_y)\sigma'_x$&$ k_xk_y\sigma'_y\otimes \bm{z}\cdot \bm{s}$&$\alpha(k^2_x-k^2_y)\sigma'_x+ k_xk_y\sigma'_y\otimes \bm{z}\cdot \bm{s}$
			\\
			\hline
			\hline
		\end{tabular}\label{tab1}	
	\end{table*}
\end{widetext}

\renewcommand*{\bibnumfmt}[1]{[S#1]}


\begin{thebibliography}{3}
\bibitem{Fn2} For example, the direct Coulomb interactions in multiorbitals models, including the Hund’s coupling, are instantaneous and are thus independent of the energy separation between any pair of electrons.
\end{thebibliography}
\end{document}